\documentclass[preprint,preprintnumbers,amsmath,amssymb]{revtex4}

\usepackage{graphicx}
\usepackage{dcolumn}
\usepackage{bm}
\begin{document}

\title{The  $p_{x}+ip_{y}$ Chiral Superconductor wire  weakly coupled to  two metallic rings pierced by  an  external flux}

\author{ D. Schmeltzer}

\affiliation{Physics Department, City College of the City University of New York \\
New York, New York 10031}

\begin{abstract}

We consider  a  p-wave   superconductor  wire coupled to two metallic rings. confined to a  one-dimensional wire. At the  two interface between the  the wire and the metallic rings  the pairing order parameter  vanishes, as result   two zero modes Majorana fermion  appear. The two metallic rings  are pierced by   external magnetic fluxes.   The special features of the  Majorana Fermions can be deduced   from the correlation between the currents in the two rings.

\end{abstract}


\maketitle
\textbf{I Introduction}

\vspace{0.2 in}

Topological  Superconductors are characterized by the invariance under  charge conjugation symmetry.
As a result of this invariance  zero modes Majorana fermion   appear at the interfaces between the superconductor   a metal.
$Sr_{2}Ru$ is the material  where Majorana fermion might be observed since the pairing order parameter is characterized by  $p_{x}+ip_{y}$  symmetry.
The electronic  excitations for the  ground  state pairing $p_{x}+ip_{y}$  are  given by  half vortices which are  zero mode  Majorana fermions \cite{Kitaev,DunghaiLee,Oreg1,Oreg2}.

The  realization of the p-wave superconductors physics and the formation of the zero mode Majorana Fermions  at the edges of the wire can be achieved using a p-wave wire of length $L$ coupled to two metallic rings which are pierced by external magnetic fields.The current in the rings is coupled to the p-wave wire trough the Majorana modes which are bound at the interface. For different fluxes in the rings we find that the excitations in the wires become imaginary resulting in an unstable vanishing current.
The only stable current are obtained for the case that the imaginary part vanishes. The stable current are obtained for special relations between the magnetic fluxes and the wire excitations energy.
This feature is attributed to the existence of the Majorana fermions.
We find that the correlation between  the currents in the two rings is affected by the presence of the Majorana Fermions.
Therefore the   model  introduced here can be used   for the  identification of  the  Majorana fermions.

The content of this paper is as follows: In chapter $II$ we present the model of the  the $p-wave$ wire coupled to two metallic rings. Using the left and right mover we obtain the continuum representation of the superconductor wire. At the two edges of the wire we obtain the two zero modes of the wire. The effect of the coupling between the wire and the metallic rings is dominated by the zero modes of the wire. As a result we derive the effective Hamiltonian between  the zero modes and the metallic rings. This effective Hamiltonian is controlled by the low energy  excitation in the wire (the coupling energy between the   two edges)  $\epsilon\approx |\Delta_{0}| e^{-L|\Delta_{0}|}$ ,$L|\Delta_{0}|>>1$ where  $\Delta_{0}$ is the pairing field and $L$ is the length of the wire. In section $III$ we consider the case that the wire energy $\epsilon\rightarrow 0$.
In section $IV$ we consider the case where the wire energy is finite.
Section $V$ is devoted to conclusions.

\vspace{0.2 in} 

\textbf{II- The model for the p-wave wire weakly coupled to two rings pierced by  external fluxes }

\vspace{0.2 in}

The  p-wave  wire  of length $L$   is given by the Hamiltonian  $H_{P-W}$ :
\begin{equation}
H_{P-W}=
-t\sum_{x,x'}(C^{+}(x)C(x')+h.c.) -\mu_{F}\sum_{x}C^{+}(x)C(x)
-\hat{\Delta}\sum_{x,x'}[\gamma_{x,x'}C^{+}(x)C^{+}(x')+\gamma_{x,x'}C(x')C(x)]
\label{pwave}
\end{equation}
The pairing gap  is given by $\hat{\Delta}$ and the  polarized  fermion operator is given by  $C(x)\equiv C_{\sigma=\uparrow}(x)$.
The matrix elements $\gamma_{x,x'} $ obey obey the p-wave symmetry: $\gamma_{x,x'=x-a}=-\gamma_{x,x'=x+a}$, $|\gamma_{x,x'}|=1$, therefore 
the time reversal  and parity symmetry  are  both  broken and  obeys the pairing  boundary conditions $\hat{\Delta}(x=0)= \hat{\Delta}(x=L)=0$ and $\hat{\Delta}(x)=\hat{\Delta}_{0}$ for $0<x<L$.

We introduce the right and left fermions in the continuum representation  for the fermions in the wire : $C(x=na)\sqrt{a}\rightarrow C(x)=e^{ik_{F}x}\hat{C}_{R}(x)+ e^{-ik_{F}x}\hat{C}_{L}(x)$ and find that equation $1$ is replaced by the Hamiltonian : 
\begin{equation}
H_{P-W}=\int dx [v_{F}\Psi^{\dagger}(x)\sigma^{z}(-i\partial_{x})\Psi(x)+ \Delta(x)\Psi^{\dagger}(x)\sigma^{x}\Psi(x)]
\label{ham}
\end{equation}  
This Hamiltonian is invariant under the charge conjugation symmetry. The  spinor  $\Psi^{\dagger}(x)=[C^{\dagger}_{R}(x),C_{L}(x)]\equiv [e^{\frac{i\pi}{4}}\hat{C}^{\dagger}_{R}(x),e^{\frac{-i\pi}{4}}\hat{C}_{L}(x)]$  satisfies the reality constraint condition $K \Psi^{\dagger}(x)= \Psi(x)$, where $K$ is the charge conjugation operator.

The pairing field  $\Delta(x)\equiv 4\hat{\Delta}(x) \sin(k_{F}a)$ can be written  as $\Delta(x)=M_{L}(x)+M_{R}(x-L)$ where   $M_{L}(x)=\frac{\Delta(0)}{2}sgn(x)$    and   $M_{R}(x-L)=\frac{\Delta(0)}{2}sgn(x-L)$ obeys the domain wall property:   $M_{L}(-x)=-M_{L}(x)$ (at x=0) , $M_{R}(-(x-L))=- M_{R}(x-L)$ (at x=L) .
The zero modes eigenfunctions  are  given by $\eta_{\lambda}(x)=[\eta_{1}(x),\eta_{2}(x)]^{T}$  and eigenstates  of the operator  $\sigma^{y}\eta_{\lambda}(x)=\lambda \eta_{\lambda}(x)$ with  $\lambda=\pm 1$.
The zero mode spinor which is localized around $x=0$  is  identified with  $\lambda=-1$
and the second one which localized around $x=L$ is identified  with $\lambda=1$
\begin{eqnarray} 
 &&\eta_{Left}(x)\equiv  \eta_{\lambda=-1}(x)=e^{\frac{-1}{v_{F}}\int_{0}^{x}\Delta(x')\,dx' } \frac{e^{i\frac{\pi}{4}}}{\sqrt{2}}[1,-i]^{T}\nonumber\\&&
\eta_{Right}(x)\equiv  \eta_{\lambda=1}(x)=e^{\frac{1}{v_{F}}\int_{L}^{x}\Delta(x')\,dx' } \frac{e^{-i\frac{\pi}{4}}}{\sqrt{2}}[1,i]^{T}\nonumber\\&&
\end{eqnarray}
The spinor operator  $\Psi(x)$ with  the  two zero mode  Majorana operators $\alpha_{l}$ (at the left edge)  and  $\alpha_{r}$  (at the right edge),  $(\alpha_{r})^2=(\alpha_{l})^2 =\frac{1}{2}$ takes the form :
\begin{equation}
\Psi(x)\rightarrow \Psi(x) + \alpha_{r}\eta_{\lambda=1}(x)
+\alpha_{l}\eta_{\lambda=-1}(x)
\label{psi}
\end{equation}
As a result the low energy of the p-wave wire  is given by:
\begin{equation}
H_{P-W}=\int dx [v_{F}\Psi^{\dagger}(x)\sigma^{z}(-i\partial_{x})\Psi(x) + \Delta(x)\Psi^{\dagger}(x)\sigma^{x}\Psi(x)]\approx  \frac{i}{2}\epsilon \alpha_{l} \alpha_{r}
\label{eqnew}
\end{equation}
where $\epsilon\approx |\Delta_{0}| e^{-L|\Delta_{0}|}$ ,$L|\Delta_{0}|>>1$.

At this stage we include the two rings Hamiltonian pierced by the fluxes $\hat{\varphi}_{i}$ , $i=1,2$ and length  $l_{ring}<<L$. The left ring is restricted to the region $-l_{ring}\leq x \leq 0$ and the right ring is restricted to  $L\leq x \leq L+l_{ring}$. Since only the wire fields at $x=0$  and  $x=L$ are involved   we  fold  the space of the  right ring $i=2$  such that both rings are restricted to the region  $-l_{ring}\leq x \leq 0$.  As a result the external fluxes obey:  $\hat{\varphi}_{1}\rightarrow \hat{\varphi}_{1}$ and  $\hat{\varphi}_{2}\rightarrow -\hat{\varphi}_{2}$.
In addition we replace for each ring the fermion operator $\psi_{i}(x)$,i=1,2 by  the $right$ $R_{i}(x)$ and $left$ fermions $L_{i}(x)$:
\begin{equation}
\psi_{i}(x)=R_{i}(x)e^{ik_{F}x}+L_{i}(x) e^{-ik_{F}x}
\label{rings}
\end{equation}  
We replace for each ring the $right$ and $left$ movers   by  four Mayorana  operators: 
\begin{eqnarray}
&&r_{i}\equiv(R_{i}(0)-R^{\dagger}_{i}(0))(-i)\nonumber\\&& l_{i}\equiv L_{i}(0)+L^{\dagger}_{i}(0)\nonumber\\&&
\end{eqnarray}
The the matrix element between the   wire and rings  is given by $-g$. As a result   the  low energy hopping Hamiltonian is given by: 
\begin{equation}
H_{T} =\frac{-ig}{\sqrt{2}}[\alpha_{l}(r_{1}+l_{1})+\alpha_{r}(r_{2}-l_{2})]
\label{majorana}
\end{equation}
We replace  the Majorana zero modes $\alpha_{l}$ and $\alpha_{r}$ by   the  fermion pair, $q=\alpha_{l}+i \alpha_{r}$,  $q^{\dagger}=\alpha_{l}-i \alpha_{r}$  which  obey: $[q,q^{\dagger}]_{+}=1$,  $q^{\dagger}|0>=|1>$ and  $q|1>=|0>$ where $|0>$  is  ground state of  wire and rings:
  $R_{i,p}(x)|0>=L_{i,p}(x)|0>= R^{\dagger}_{i,h}(x)|0>=L^{\dagger}_{i,h}(x)|0>= q|0>=0$. Where   $ R_{i,p}(x)$, $L_{i,p}(x)$  are the particle operators  and   $R^{\dagger}_{i,h}(x)$, $L^{\dagger}_{i,h}(x)$ are the holes operators $R^{\dagger}_{i,h}(x)$, $L^{\dagger}_{i,h}(x)$.  The right and left mover are given as a linear combination of particles  and holes operators. $ R_{i,p}(x)$ $L_{i,p}(x)$ represent the annihilation of particles  and   $ R^{\dagger}_{i,h}(x)$, $L^{\dagger}_{i,h}(x)$  are the creation operators for holes.
\begin{equation}
R_{i}(x)= R_{i,p}(x)+R^{\dagger}_{i,h}(x);\hspace{0.1 in} L_{i}(x)= L_{i,p}(x)+L^{\dagger}_{i,h}(x)
\label{operators}
\end{equation} 
Using the Fermionic representation  we replace  $H_{P-W}$  given in equation $5$  and $H_{T}$ given in equation  $(8)$ by :
\begin{equation}
H_{P-W}+H_{T}\equiv\epsilon q^{\dagger}q -\frac{ig}{2\sqrt{2}}[(q+q^{\dagger})(r_{1}+l_{1}) -i(q-q^{\dagger})(r_{2}-l_{2})]
\label{energy}
\end{equation}
The value of the wire  energy $\epsilon$ in equations $5,10$ is based on the projection of the spinor (in equation $4$)  on  the zero modes  $\eta_{\lambda=1}(x)$ and 
$\eta_{\lambda=-1}(x)$.  Since the leads couple to the modes in the wire, we expect that the non-zero modes will give rise to a finite width of the energy $\epsilon$.  Therefore for finite energies we will replace  $\epsilon$ by $\hat{\epsilon}\equiv  \epsilon -i\Gamma$, where the width $\Gamma \propto  g^4$.

 We perform an exact integration over the  fermion operators  $q$,$q^{\dagger}$ and find the time dependent effective interaction  $H_{eff}(t)$:
\begin{eqnarray}
&&H_{eff}(t)=\frac{-i g^2}{2} \int\,dt'\mu[t,t']e^{-i \frac{\hat{\epsilon}}{\hbar}(t-t')}[r_{2}(t)-l_{2}(t)+i(r_{1}(t)+l_{1}(t))][r_{2}(t')-l_{2}(t')-i(r_{1}(t')+l_{1}(t'))]\nonumber\\&&
\end{eqnarray}
where $\mu[t,t']$ is the step function which is one for $t>t'$.

\vspace{0.2 in}

\textbf{III-The  effective interaction  $H_{eff}(t)$ in the limit  $\epsilon\rightarrow0$}

\vspace{0.2 in}

When $\epsilon\rightarrow0 $  equation $11$ is replaced by :

\begin{eqnarray}
&&H_{eff}(t)=\frac{-ig^2}{2} \int\,dt'\mu[t,t'][r_{2}(t)-l_{2}(t)+i(r_{1}(t))+l_{1}(t)][r_{2}(t')-l_{2}(t')-i(r_{1}(t')+l_{1}(t')]\nonumber\\&&
\end{eqnarray}
Using  the scaling analysis given  in \cite{davidimpurity} we observe that  the effective interaction   flows to the strong coupling limit  and find  $g^{2}(b)=g^{2}b^{2-\alpha}$,  $\alpha\approx 1$ where  $b>1 $. Since the coupling constant   $g(b)$ flows to infinity, the only way  a  solution  will exists if the effective interaction annihilates the ground state:  $H_{eff}(t)|0>=0$.
 Therefore the  physical solution  is given by the $constraint$   \cite{rings} equation: 
\begin{equation}
[r_{2}(t)-l_{2}(t)-i(r_{1}(t)+l_{1}(t)]|0>=0
\label{c}
\end{equation}  
Since  $R_{i,p}(x)|0>=L_{i,p}(x)|0>= R^{\dagger}_{i,h}(x)|0>=L^{\dagger}_{i,h}(x)|0>=0$ the constraint condition implies for particles  ($p$ stands for particles)
the equation  :  
$[i(R^{\dagger}_{2,p}-L^{\dagger}_{1,p})-(L^{\dagger}_{2,p}-R^{\dagger}_{1,p})]|0>=0$;
and  for holes  ($h$ stands for holes)
$[i(R_{2,h}-L_{1,h})-(L_{2,h}-R_{1,h})]|0>=0$.
We find the constraint equation:
\begin{equation}
\psi_{1}(x=0)\equiv R_{1}(x=0)+L_{1}(x=0)=e^{-i\frac{\pi}{2}}[R_{2}(x=0)+L_{2}(x=0)]\equiv e^{-i\frac{\pi}{2}}\psi_{2}(x=0)=\widetilde{\psi}_{2}(x=0)
\label{eqco}
\end{equation}
The explicit identity contains the phase factor  $e^{-i\frac{\pi}{2}}$  which is obtained from Bosonization (see below).
Following equation $14$ we find  the constraint condition  for the ground state $|0>$:
\begin{equation}
\kappa\equiv [\psi_{1}(x)-\psi_{2}(x)]|_{x=0},\hspace{0.2 in}  \kappa|0>=0 
\label{equation}
\end{equation}
Following \cite{rings} we find two additional constraints  equation :
\begin{equation}
\mathcal{E}\equiv[(-i\partial_{x}- \frac{2\pi}{l_{ring}}\hat{\varphi}_{1})^{2}\psi_{1}(x)- (-i\partial_{x}+ \frac{2\pi}{l_{ring}}\hat{\varphi}_{2})^{2}\psi_{2}(x)]|_{x=0}, \hspace{0.2 in} \mathcal{E}|0>=0
\label{energy}
\end{equation}
\begin{equation}
\mathcal{J}\equiv [(-i\partial_{x}- \frac{2\pi}{l_{ring}}\hat{\varphi}_{1})\psi_{1}(x)+(-i\partial_{x}+ \frac{2\pi}{l_{ring}}\hat{\varphi}_{2})\psi_{2}(x)]|_{x=0},\hspace{0.2 in} \mathcal{J}|0>=0
\label{j}
\end{equation}
Any eigenstate of N particles must satisfy the set of equations  $15-17$  with the periodic boundary condition $\psi_{i}(x)=\psi_{i}(x+l_{ring})$. For the case $N=1$ (one particle) we have :
\begin{equation}
|N=1>=\int_{-l_{ring}}^{0}\,dx[f_{1}(x)\psi^{\dagger}_{i}(x)+f_{2}(x)\psi^{\dagger}_{2}(x)]
\label{solone}
\end{equation}
 where $f_{i}(x)$ are the amplitudes . Using  equations  $15-17$  we  find finite  solutions for the amplitudes  $f_{i}(x)$  only when the fluxes are equal. 

\vspace{0.2 in}

\textbf{IV-The finite  limit $\epsilon\neq0$}

\vspace{0.2 in}

In order to study the finite limit $\epsilon\neq0$ we will use the zero mode Bosonization method  \cite{davidimpurity,Berkovits}. The right $R_{i}(x)$ and left $ L_{i}(x)$ fermions  for each ring $i=1,2$ is given by:
\begin{eqnarray}
&&R_{i}(x)=\sqrt{\frac{\Lambda}{2\pi}}Z_{i}e^{i\alpha_{R,i}}e^{\frac{2\pi}{l_{ring}}(N_{R,i}-\frac{1}{2})x}e^{i\sqrt{4\pi}\vartheta_{R,i}(x)}\nonumber\\&&
L_{i}(x)=\sqrt{\frac{\Lambda}{2\pi}}Z_{i}e^{i\alpha_{L,i}}e^{\frac{2\pi}{l_{ring}}(N_{L,i}-\frac{1}{2})x}e^{i\sqrt{4\pi}\vartheta_{L,i}(x)}\nonumber\\&&
\end{eqnarray}
Where $Z_{1}Z_{2}=-Z_{2}Z_{1}$ are Majorana  variable which ensure the anti-commutation between  the two rings in the bosonic representation.
In the bosonic representation we have the zero modes  $\alpha_{R,i}$  $\alpha_{L,i}$  bosons and their conjugates ,$N_{R,j}$,$N_{L,j}$.  The zero modes obey  the commutation rules :$ [-\alpha_{L,i},N_{L,j}]=i\delta_{i,j}$ and 
$[\alpha_{R,i},N_{R,j}]=i\delta_{i,j}$.

As a result we obtain the zero mode representation  in terms of the  fermion numbers  $N_{L,i}$,$N_{R,i}$ and fluxes  $\hat{\varphi}_{i}$ in each ring: 
\begin{eqnarray}
&&H_{0}=\frac{\pi v_{F} \hbar}{2 l_{ring}}[(N_{L,1}-N_{R,1}+2\hat{\varphi}_{1})^2+(N_{L,1}+N_{R,1})^2] +\frac{\pi v_{F} \hbar}{2 l_{ring}}[(N_{L,2}-N_{R,2}+2\hat{\varphi}_{2})^2+(N_{L,1}+N_{R,1})^2]\nonumber\\&&
\end{eqnarray}
Using the equations of motion $i\hbar\frac{d \alpha_{R,i}}{dt}=[\alpha_{R,i},H_{0}]$ and  $i\hbar\frac{d \alpha_{L,i}}{dt}=[\alpha_{L,i},H_{0}]$ ,$i=1,2$ we obtain   the zero mode representation in the \textbf{interaction picture}.
We will use the zero mode representation  $\alpha^{I}_{R,i}(t)$, $\alpha^{I}_{L,i}(t)$ in the \textbf{interaction picture} in order  to evaluate  equation  $15$. We find that  $H_{eff}(t)$ is given in terms of the zero mode functions  $F(t+\frac{\tau}{2})$ and $G(t-\frac{\tau}{2})$ :
\begin{equation}
H_{eff}(t)=-2i\hat{g}^2\int_{0}^{\infty} \,d \tau F(t+\frac{\tau}{2})e^{-i\frac{\hat{\epsilon}}{\hbar}\tau}G(t-\frac{\tau}{2})
\label{eqt}
\end{equation}
where $\hat{g}^2=g^2\frac{\Lambda}{2\pi}$ is the  coupling  constant and   $\nu_{0}\equiv \frac{\epsilon}{\hbar}$, $\nu_{i}\equiv\frac{2\pi v_{F}}{l_{ring}}\hat{\varphi}_{i}$ are the equivalent wire and rings frequencies and $\hat{\Gamma}=\frac{\Gamma}{\hbar} $ is the width. 
The  functions  $F(t+\frac{\tau}{2})$ and $G(t-\frac{\tau}{2})$  are given by:
\begin{eqnarray}
&&F(t+\frac{\tau}{2})=Z_{2}((\sin\alpha^{I}_{R,2}(t+\frac{\tau}{2})-\cos\alpha^{I}_{L,2}(t+\frac{\tau}{2}))+iZ_{1}((\sin\alpha^{I}_{R,1}(t+\frac{\tau}{2})-\cos\alpha^{I}_{L,1}(t+\frac{\tau}{2}))=\nonumber\\&&Z_{2}[\sin\alpha^{I}_{R,2}(t)\cos(\nu_{2}\tau)+ \cos\alpha^{I}_{R,2}(t)\sin(\nu_{2}\tau)-\cos\alpha^{I}_{L,2}(t)\cos(\nu_{2}\tau)+\sin\alpha^{I}_{L,2}(t)\sin(\nu_{2}\tau)]\nonumber\\&&+
iZ_{1}[\sin\alpha^{I}_{R,1}(t)\cos(\nu_{1}\tau)- \cos\alpha^{I}_{R,1}(t)\sin(\nu_{1}\tau)+\cos\alpha^{I}_{L,1}(t)\cos(\nu_{1}\tau)-\sin\alpha^{I}_{L,1}(t)\sin(\nu_{1}\tau)]\nonumber\\&&
\end{eqnarray}
\begin{eqnarray}
&&G(t-\frac{\tau}{2})=Z_{2}((\sin\alpha^{I}_{R,2}(t-\frac{\tau}{2})-\cos\alpha^{I}_{L,2}(t-\frac{\tau}{2}))-iZ_{1}((\sin\alpha^{I}_{R,1}(t-\frac{\tau}{2})+\cos\alpha^{I}_{L,1}(t-\frac{\tau}{2}))=\nonumber\\&&Z_{2}[\sin\alpha^{I}_{R,2}(t)\cos(\nu_{2}\tau)- \cos\alpha^{I}_{R,2}(t)\sin(\nu_{2}\tau)-\cos\alpha^{I}_{L,2}(t)\cos(\nu_{2}\tau)-\sin\alpha^{I}_{L,2}(t)\sin(\nu_{2}\tau)]\nonumber\\&&-i
Z_{1}[\sin\alpha^{I}_{R,1}(t)\cos(\nu_{1}\tau)+ \cos\alpha^{I}_{R,1}(t)\sin(\nu_{1}\tau)+\cos\alpha^{I}_{L,1}(t)\cos(\nu_{1}\tau)-\sin\alpha^{I}_{L,1}(t)\sin(\nu_{1}\tau)]\nonumber\\&&
\end{eqnarray}
We perform the integration with respect $\tau$ and find:
\begin{equation}
H_{eff}\approx H^{real}_{eff}+i H^{Im.}_{eff}
\label{effective}
\end{equation}
where $H^{real}_{eff}$ is the real part of the effective action:
\begin{eqnarray}
&&H^{real}_{eff}=\nonumber\\&&\frac{\hbar\hat{g}^2 \nu_{0}}{\nu_{0}^2 +\hat{\Gamma}^2}[\cos(2\alpha_{R,2})-\cos(2\alpha_{L,2})-\cos(2\alpha_{L,1})+\cos(2\alpha_{R,1})+2(\sin(\alpha_{R,2}+\alpha_{L,2})-\sin(\alpha_{R,1}+\alpha_{L,1}))]\nonumber\\&&-2\hbar\hat{g}^2 (\frac{\nu_{0}-\nu_{1}}{(\nu_{0}-\nu_{1})^2+\hat{\Gamma}^2}+\frac{\nu_{0}+\nu_{1}}{(\nu_{0}+\nu_{1})^2+\hat{\Gamma}^2})[1+\sin(\alpha_{R,1}-\alpha_{L,1})]\nonumber\\&&-2\hbar\hat{g}^2 (\frac{\nu_{0}-\nu_{2}}{(\nu_{0}-\nu_{2})^2+\hat{\Gamma}^2}+\frac{\nu_{0}+\nu_{2}}{(\nu_{0}+\nu_{2})^2+\hat{\Gamma}^2})
[1+\sin(\alpha_{R,2}-\alpha_{L,2})]\nonumber\\&&
\end{eqnarray}
The imaginary part  $H^{Im.}_{eff}$ of the action causes the current to vanish.  Finite solutions will be obtained for the  ground states $|0>$ which  obey $H^{Im.}_{eff}|0>=0$. Therefore the solutions  for   finite currents   are equivalent to a constraint condition for the ground state  $|0>$.
\begin{eqnarray}
&&H^{Im.}_{eff}=Z_{2}Z_{1}\hat{g}^2[\frac{ \hat{\Gamma}}{(\nu_{0}-\frac{\nu_{2}-\nu_{1}}{2})^2 + \hat{\Gamma}^2}+ \frac{\hat{\Gamma}}{(\nu_{0}+\frac{\nu_{2}-\nu_{1}}{2})^2+ \hat{\Gamma}^2}]\nonumber\\&&[(\sin(\alpha_{R,2})-\cos(\alpha_{L,2}))(\cos(\alpha_{L,1})+\sin(\alpha_{R,1}))+(\sin(\alpha_{R,2})+\sin(\alpha_{L,2}))\nonumber\\&&(\cos(\alpha_{R,1})+\sin(\alpha_{L,1}))]\nonumber\\&& + Z_{2}Z_{1}\hat{g}^2[\frac{\hat{\Gamma}}{(\nu_{0}-\frac{\nu_{2}+\nu_{1}}{2})^2+\hat{\Gamma}^2}+ \frac{\hat{\Gamma}}{(\nu_{0}+\frac{\nu_{2}+\nu_{1}}{2})^2+\hat{\Gamma}^2}]\nonumber\\&&[(\sin(\alpha_{R,2})-\cos(\alpha_{L,2}))(\cos(\alpha_{L,1})+\sin(\alpha_{R,1}))-(\cos(\alpha_{R,2})+\sin(\alpha_{L,2}))\nonumber\\&&(\cos(\alpha_{R,1})-\sin(\alpha_{L,1}))]\nonumber\\&&
\end{eqnarray}
We find  two cases for which  a finite solution exist.

The first case corresponds to   $\nu_{0}\approx (|\frac{\nu_{2}-\nu_{1}}{2}|)$  with the solution:
\begin{eqnarray}
&&H^{Im.}_{eff}|0>=0\nonumber\\&&
\alpha_{R,2}= \alpha_{R,1}+u\nonumber\\&&
\alpha_{L,2}= \alpha_{L,1}-u\nonumber\\&&
\end{eqnarray}
The second case  corresponds to  $\nu_{0}\approx (\frac{\nu_{2}+\nu_{1}}{2})$ with the solution:
\begin{eqnarray}
&&H^{Im.}_{eff}|0>=0\nonumber\\&&
\alpha_{R,2}= -\alpha_{R,1}+u\nonumber\\&&
\alpha_{L,2}=- \alpha_{L,1}-u\nonumber\\&&
\end{eqnarray}
We introduce the definitions for the  zero mode fields :
\begin{equation}
\alpha_{1}=\alpha_{2}\equiv \alpha ;\hspace{0.2 in}  \beta_{2}=\beta_{1}-2u,
\beta_{1}\equiv \beta
\label{eqc}
\end{equation}
The solutions  are independent on the arbitrary field  $u$ which  plays the role of a gauge condition and has to be integrated out. We integrated over the field $u$ we  find the  \textbf{effective Hamiltonian for the conditions $\nu_{0}\approx (\frac{\nu_{2}-\nu_{1}}{2})$ and   $\nu_{0}\approx (\frac{\nu_{2}+\nu_{1}}{2})$}.
We introduce the magnetic flux  the  variables  $\bar{\varphi}$ and  $\Delta$: 
\begin{equation}
\bar{\varphi}= \frac{\hat{\varphi}_{1}+\hat{\varphi}_{2}}{2};
\Delta=\frac{\hat{\varphi}_{1}-\hat{\varphi}_{2}}{2}
\label{results}
\end{equation}
We can write   both cases in a closed form:
\begin{eqnarray}
&&\frac{H}{\hbar}\approx\frac{\pi v_{F} }{2 l_{ring}}[(-i\frac{d}{d \alpha}+2\hat{\varphi}_{1})^2+(-i\frac{d}{d \beta})^2] +\frac{\pi v_{F}}{2 l_{ring}}[(-i\frac{d}{d \alpha}+2\hat{\varphi}_{2})^2+(-i\frac{d}{d \beta})^2]+\nonumber\\&&
[\frac{2\hat{g}^2 \nu _{0}}{\nu_{0}^2+ \hat{\Gamma}^2}\sin(\alpha)\sin(\beta)
-2\hat{g}^2[ \frac{\nu_{0}-(\bar{\varphi}+\Delta)}{(\nu_{0}-(\bar{\varphi}+ \Delta)^2+\hat{\Gamma}^2}+ \frac{\nu_{0}+(\bar{\varphi}+\Delta)}{(\nu_{0}+(\bar{\varphi} +\Delta)^2+\hat{\Gamma}^2}]\sin (\beta)\nonumber\\&&
-2\hat{g}^2[ \frac{\nu_{0}-(\bar{\varphi}+\Delta)}{(\nu_{0}-(\bar{\varphi}+ \Delta)^2+\hat{\Gamma}^2}+ \frac{\nu_{0}+(\bar{\varphi}+\Delta)}{(\nu_{0}+(\bar{\varphi} +\Delta)^2+\hat{\Gamma}^2}+\frac{\nu_{0}-(\bar{\varphi}-\Delta)}{(\nu_{0}-(\bar{\varphi}- \Delta)^2+\hat{\Gamma}^2}+ \frac{\nu_{0}+(\bar{\varphi}-\Delta)}{(\nu_{0}+(\bar{\varphi} -\Delta)^2+\hat{\Gamma}^2}]]\nonumber\\&&
(\delta_{\frac{\hat{\varphi}_{1}-\hat{\varphi}_{2}}{2},\nu_{0}}+\delta_{\frac{\hat{\varphi}_{1}+\hat{\varphi}_{2}}{2},\nu_{0}})
\nonumber\\&&
\end{eqnarray}
The first line of equation $31$ represents the the Hamiltonian for the two metallic rings pierced by the external fluxes expressed in therms of the zero mode of the metallic rings.
The second part of equation $31$ represents the coupling between the wire and the two rings.
We observe that this part is restricted by the constraint condition $\frac{\hat{\varphi}_{1}-\hat{\varphi}_{2}}{2}=\nu_{0}$  or  $\frac{\hat{\varphi}_{1}+\hat{\varphi}_{2}}{2}=\nu_{0}$. This constraints represents the effect of the Majorana fermions on the p-wave wire.

In order two investigate the Hamiltonian in equation $31$ we will use the algebra of the zero modes  \cite{david,davidDirac,Berkovits}: 
\begin{equation}
\hat{J}=N_{R}-N_{L}\equiv -2i\frac{d}{d \alpha}; \hspace{0.2 in}\hat{Q}=N_{R}+N_{L}\equiv -2i\frac{d}{d \beta} 
\label{eqj}
\end{equation}
with the eigenvalues and commutation rules:
\begin{eqnarray}
&&\hat{J}|J,Q>=J|J,Q> ;J=0,\pm1,\pm2,..\nonumber\\&&  \hat{Q}|J,Q>=Q|J,Q> ;Q=0,\pm1,\pm2,..\nonumber\\&&
\end{eqnarray}
From the commutation relations $ [\alpha,\hat{J}]=2i$, $ [\beta,\hat{Q}]=2i$ we establish the relations :
\begin{eqnarray}
&&e^{i\alpha}|J,Q>=|J+1,Q>  ; e^{-i\alpha}|J,Q>=|J-1,Q> \nonumber\\&&  e^{i\beta}|J,Q>=|J,Q+1> ; e^{-i\beta}|J,Q>=|J,Q-1>\nonumber\\&&
\end{eqnarray}
The eigenfunctions are given by:
\begin{eqnarray}  
&&<\alpha|J,Q=0>=\frac{1}{\sqrt{4\pi}}e^{i\alpha J} ; <\beta|J=0,Q>=\frac{1}{\sqrt{4\pi}}e^{i\alpha J}\nonumber\\&&
\end{eqnarray}

Using  the algebra of the zero modes we compute to  lowest order (in perturbation theory) the energy for the ground state of the two rings coupled to the wire. As a function of the coupling constant  $\lambda\equiv\frac{2\hat{g}^2}{\nu_{max}}<1$  and  maximum frequency  $\nu_{max}$  which is given by  the electronic bandwidth frequency. We find for the ground state energy  $E(\hat{\varphi}_{1},\hat{\varphi}_{2})$:
\begin{eqnarray}
&&E(\hat{\varphi}_{1},\hat{\varphi}_{2})=[\frac{2\hbar\pi v_{F} }{ l_{ring}}((\bar{\varphi}+\Delta)^2+ (\bar{\varphi}-\Delta)^2)\nonumber\\&&
-\lambda(\frac{\nu_{max}}{\nu_{0}})( \frac{\nu_{0}-(\bar{\varphi}+\Delta)}{(\nu_{0}-(\bar{\varphi}+ \Delta)^2+\hat{\Gamma}^2}+ \frac{\nu_{0}+(\bar{\varphi}+\Delta)}{(\nu_{0}+(\bar{\varphi} +\Delta)^2+\hat{\Gamma}^2}+\nonumber\\&&\frac{\nu_{0}-(\bar{\varphi}-\Delta)}{(\nu_{0}-(\bar{\varphi}- \Delta)^2+\hat{\Gamma}^2}+ \frac{\nu_{0}+(\bar{\varphi}-\Delta)}{(\nu_{0}+(\bar{\varphi} -\Delta)^2+\hat{\Gamma}^2})](\delta _{\nu_{0},\bar{\varphi}}+\delta _{\nu_{0},\Delta})
\nonumber\\&&
\end{eqnarray}
Using eq.36 we  compute   the currents  $I_{i}=\frac{\partial  E(\hat{\varphi}_{1},\hat{\varphi}_{2})}{\partial \hat{\varphi}_{i}}$ for the two rings  $i=1,2$ using the conditions:
$\nu_{0}\approx \frac{\hat{\varphi}_{1}-\hat{\varphi}_{2}}{2}\equiv \Delta$. 
 
 
The current in ring one $I_{1}$ and ring two $I_{2}$ are represented  in terms  of 
$\bar{\varphi}= \frac{\hat{\varphi}_{1}+\hat{\varphi}_{2}}{2}$,  $\nu_{0}= \Delta$ and 
the current amplitude $I_{0}=\frac{2\pi v_{F}}{l_{ring}}$
 
\begin{eqnarray}
&&\frac{I_{1}}{I_{0}}=\bar{\varphi}+\nu_{0}-\lambda[\frac{\bar{\varphi}^2-\hat{\Gamma}^2}{(\bar{\varphi}^2+\hat{\Gamma}^2)^2}+\frac{-(2\nu_{0}+\bar{\varphi})^2+\hat{\Gamma}^2}{(2\nu_{0}+\bar{\varphi})^2+\hat{\Gamma}^2)^2}]\nonumber\\&&
\frac{I_{2}}{I_{0}}=\bar{\varphi}-\nu_{0}-\lambda[\frac{\bar{\varphi}^2-\hat{\Gamma}^2}{(\bar{\varphi}^2+\hat{\Gamma}^2)^2}+\frac{-(2\nu_{0}+\bar{\varphi})^2+\hat{\Gamma}^2}{(2\nu_{0}+\bar{\varphi})^2+\hat{\Gamma}^2)^2}]\nonumber\\&&
\frac{I_{1}+I_{2}}{2 I_{0}}=[\bar{\varphi}-\lambda(\frac{\bar{\varphi}^2-\hat{\Gamma}^2}{(\bar{\varphi}^2+\hat{\Gamma}^2)^2}+\frac{-(2\nu_{0}+\bar{\varphi})^2+\hat{\Gamma}^2}{(2\nu_{0}+\bar{\varphi})^2+\hat{\Gamma}^2)^2})]\nonumber\\&&
\end{eqnarray}
In figure $1$ we have plotted the  current $ \frac{I_{1}+I_{2}}{2 I_{0}}$ as a function of the flux $\bar{\varphi}$ for the case $\nu_{0}\approx \frac{\hat{\varphi}_{1}-\hat{\varphi}_{2}}{2}\equiv \Delta=0.01$. We observe that for $\bar{\varphi}>0.1$ the current is proportional to $\bar{\varphi}$. From other-hand when the $\bar{\varphi}<0.1$  the current in each ring is affected by the flux in the other ring.  This is seen  from the negative contribution of the current shown in figure $1$.The negative current contribution might be related to the Andreev reflection which occurs at the interfaces between the superconductor and the metal.

In figure $2$ we have plotted the  current $ \frac{I_{1}+I_{2}}{2 I_{0}}$ as a function of the flux $\bar{\varphi}$ for the case  $\epsilon =0$ considered  in chapter $III$. Due to the  constraint condition $\nu_{0}= \frac{\hat{\varphi}_{1}-\hat{\varphi}_{2}}{2}$  we have  the relation $\nu_{0}= \frac{\hat{\varphi}_{1}-\hat{\varphi}_{2}}{2}=\Delta=0.$. We find   a stable current in agreement with chapter  $III$ where  the current scales linearly with the flux $\hat{\varphi}_{1}=\hat{\varphi}_{2}=\bar{\varphi}$.

\vspace{0.2 in}

\textbf{V-Conclusion}

\vspace{0.2 in}

We  have investigate the dependence of the current on   the fluxes  for the entire   regime of parameters   \cite{davidimpurity,berkovits}
In  the limit of large $L$ , $\epsilon \rightarrow 0$  the current vanishes in both  rings when the two fluxes are different.  We observe   that for a finite energy  $\epsilon $  and different fluxes   the current dependence is more complex. When the two fluxes are almost equal  the current  is a function of the averaged flux. For the case that the flux difference is comparable to the flux average, the current changes sign. We can interpret this effect as  an Andreev reflection and represents a finger print of the Majorana fermions.

\pagebreak

\clearpage
\begin{figure}
\begin{center}
\includegraphics[width=6.5 in ]{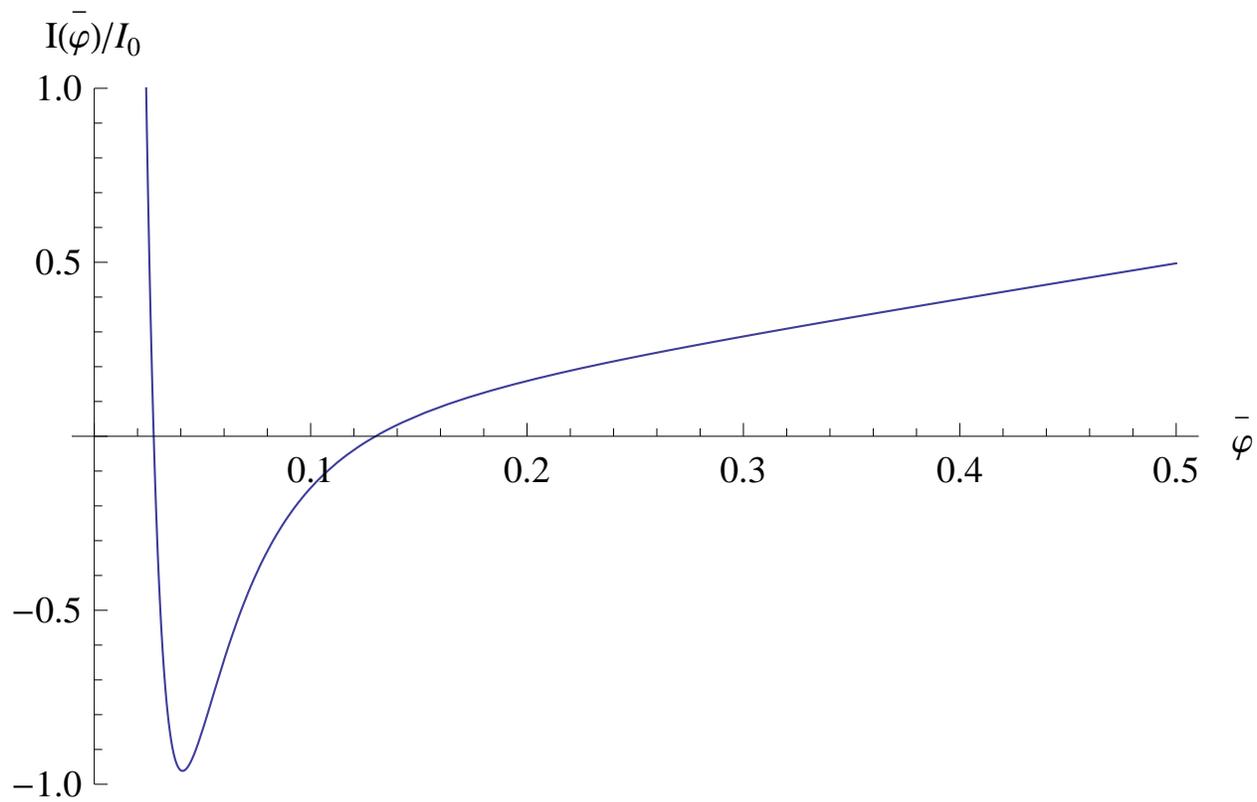}
\end{center}
\caption{The average current as a function of $\bar{\varphi}$  for the condition  $\nu_{0}\approx \frac{\hat{\varphi}_{1}-\hat{\varphi}_{2}}{2}\equiv \Delta=0.01$}  
\end{figure}

\pagebreak

\clearpage
\begin{figure}
\begin{center}
\includegraphics[width=6.5 in ]{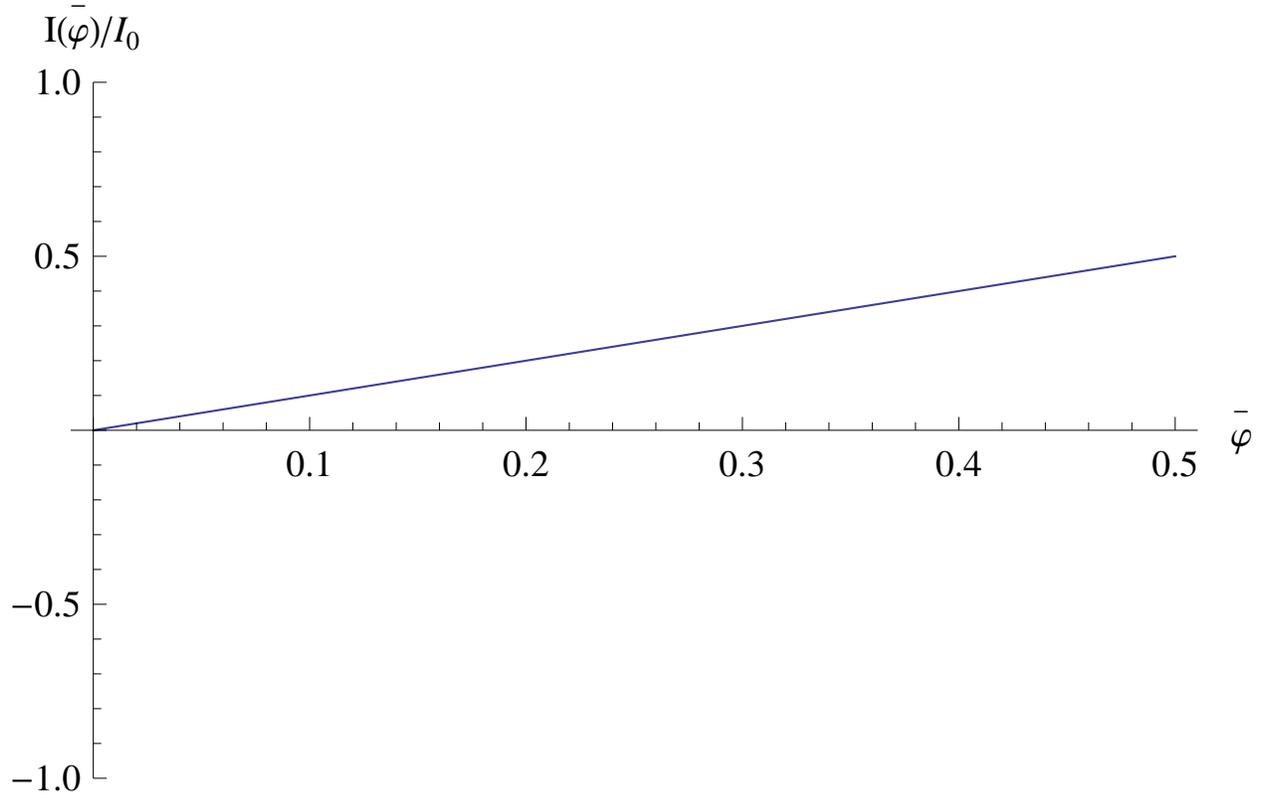}
\end{center}
\caption{The average current as a function of $\bar{\varphi}$  for the case $\epsilon=0$, $\nu_{0}=\frac{\hat{\varphi}_{1}-\hat{\varphi}_{2}}{2}\equiv \Delta=0.$}  
\end{figure}


\begin{thebibliography}{99}

\bibitem{Kitaev} Alexei Yu Kitaev "`Unpaired Majorana Fermions In Quantum Wires"'
cond-mat/ 0010440 and Alexei Yu Kitaev ,Ann.Phys.\textbf{303},2 (2003).

\bibitem{davidtop}D.Schmeltzer "`Topological Insulators"'-Transport In Curved Space"'
 Advances in Condensed Matter and Material Research Volume \textbf{10} pp. 379-402
Editors: Hans Geelvinck and Sjaak Reynst   and cond-mat/1012.5876.

\bibitem{genus} D.Schmeltzer , J. Phys:Condens Matter \textbf{20} 335205(2008). 

\bibitem{rings}  D.Schmeltzer  and A.Saxena ,Phys.Rev.B \textbf{81} ,195310 (2010).

\bibitem{DunghaiLee} S.Tewari,S.Das Sarma and Dung-Hai Lee"' An index Theorem For The Majorana Zero Modes, In Chiral P-Wave Superconductors"' cond-ma/0609556



\bibitem{Oreg1}Y.Oreg ,Gil Refaeli and Felix von Oppen "`Helical Liquids and Majorana Bound States In Quantum Wires"' cond-mat/1003.1145

\bibitem{Oreg2} L.Jiang, D.Peccker,J.Alice,Gil Refaeli, Y.Oreg and Felix von Oppen 
cond-mat/1107.4102

\bibitem{Berkovits} D.Schmeltzer and R.Berkovits Physics Letters A \textbf{253} 341-344(1999).

\bibitem{daviddirac} D.Schmeltzer "`Dirac's Method For Constraints Quantum Wires"' J.Phys:Condens.Matter \textbf{23} 155601 (2011).

 \bibitem{SemenoffSodano} Gordon W.Semenoff  and Pasquale Sodano "` Teleportation By A Majorana Medium"' cond-mat/0601261.

\bibitem{Boyanovsky}  Daniel Boyanovsky 
Phys.Rev.B.\textbf{39}, 6744(1989).

\bibitem{davidimpurity} D.Schmeltzer et al.,Phys.Rev.Lett\textbf{90},116802(2003)

\bibitem{david} D.Schmeltzer et. al  J.Phys.Condens. Matter \textbf{22}, 095301  (2010)

\bibitem{davidDirac} D.Schmeltzer   J.Phys.Condens. Matter \textbf{23},155601  (2011)

\bibitem{berkovits} D.Schmeltzer and R.Berkovits 
Phys.Lett.A\textbf{243}341,(1999)

\end{thebibliography}
\end{document}